\documentclass[11pt]{article}
\usepackage[margin=1in]{geometry}
\usepackage{amsmath,amssymb}
\usepackage{graphicx}
\usepackage{booktabs,tabularx}
\usepackage{microtype}
\usepackage[backend=biber,style=numeric-comp,sorting=none,sortcites=true,giveninits=true,maxbibnames=99]{biblatex}
\DefineBibliographyStrings{english}{andothers={\textit{et al\adddot}}}
\usepackage[hidelinks]{hyperref}
\title{Inverse design of poles and zeros: tailoring phase control in metasurfaces}
\author{Junhe Zhang\textsuperscript{1}, Seoyoung Kim\textsuperscript{2}, Hyun Jung Kim\textsuperscript{2}, Juejun Hu\textsuperscript{1,*}\\[0.5em]
\small \textsuperscript{1}Department of Materials Science and Engineering,\\
\small Massachusetts Institute of Technology,\\
\small 77 Massachusetts Avenue, Cambridge, MA 02139, USA\\[0.3em]
\small \textsuperscript{2}The Department of Aerospace Engineering, KAIST, Daejeon, South Korea\\[0.3em]
\small Email: \href{mailto:hujuejun@mit.edu}{hujuejun@mit.edu}}
\date{}

\begin{document}

\maketitle

\begin{abstract}
We develop a physically interpretable inverse-design framework for \emph{generalized} Huygens metasurfaces, using prescribed scattering poles and zeros to engineer resonance interference and transmission-phase control. Compared with direct optimization of transmission or reflection coefficients, this approach offers a physically transparent route to rationally controlling resonance properties, including linewidth and group delay, while reducing computational work. The resulting designs demonstrate how near-unity transmission and $2\pi$ phase excursion can arise from resonance interference. These signatures of the Kerker effect extend beyond the conventional alignment of electric and magnetic dipole resonances to other mechanisms, including guided-mode resonances and quasi-bound states in the continuum. We further optimize the Kerker condition across a continuum of refractive-index states in a reconfigurable phase change material metasurface, enabling high-transmission $2\pi$ phase tuning.
Our results establish a versatile and practical approach to metasurface phase control through the targeted design of scattering poles and zeros.
\end{abstract}

\section{Introduction}
Optical metasurfaces control light through the collective scattering response of patterned interfaces~\cite{Yu2014Flat}. A central objective in transmissive metasurface design is to realize full $2\pi$ phase control while maintaining near-unity transmission efficiency. A widely used approach employs waveguide-type meta-atoms that act as truncated waveguides, accumulating a predominantly nonresonant propagation phase; their geometries are often selected to avoid strong resonances within the operating band~\cite{Chen2023Dispersion,Arbabi2015}. Resonant metasurfaces offer an alternative route, exemplified by classical Huygens surfaces, in which balanced electric and magnetic responses suppress backward scattering while producing a large transmitted phase change~\cite{Pfeiffer2013,Kerker1983,Staude2013}. Overlapping electric and magnetic dipole resonances can provide high transmission and full phase coverage in low-profile structures without requiring large propagation lengths or high aspect ratios~\cite{Decker2015,Yu2015,Zhang2018,Chong2016,Zhao2016}. Beyond their reduced thickness, resonant designs offer versatile control of dispersion and group delay, wavelength-selective functionality, and enhanced light--matter interaction through resonant field buildup~\cite{Kuznetsov2016,Jahani2016}. Their compact geometry and sensitivity to material properties also make them attractive for reconfigurable metasurfaces~\cite{Gu2023,Shalaginov2021}. Refractive-index tuning shifts the resonances through a fixed operating wavelength, enabling a suitably engineered transmission response to span a full $2\pi$ phase range. For high-quality-factor resonances, the steep phase variation allows even small index perturbations to produce large phase changes. These opportunities extend beyond dipoles in isolated particles: periodic structures also support guided and collective resonances~\cite{Wang1993GMR,Hsu2016,Koshelev2018}, motivating a description expressed directly in terms of observable scattering channels rather than a particular meta-atom geometry.

Analytic continuation of transmission and reflection into the complex-frequency plane supplies such a description. Resonance poles encode oscillation frequencies and decay rates, whereas channel zeros specify destructive interference in a selected output. For metasurface phase control, Colom \textit{et al.} connected a full phase winding to the relative positions of poles and zeros and the branch structure of the complex phase~\cite{Colom2023}. Binkowski \textit{et al.} developed contour-based methods for extracting poles, zeros, and their sensitivities, and used modal expansions to resolve the origin of reflection zeros in a metasurface~\cite{Binkowski2024}. These studies and other work~\cite{Mikheeva2023,Elsawy2023,Alpeggiani2017,Lalanne2018,Kristensen2020} establish poles and zeros as physically meaningful descriptors of resonance interference and phase delay.

While these studies have broadened our understanding of resonance phenomena and phase-delay mechanisms, using tailored poles and zeros to construct generalized Huygens responses remains underexplored. Chen \textit{et al.} established an important methodological foundation: an adjoint-based framework for multiresonance filter design that enforces prescribed complex poles and port-coupling ratios through conjugate-frequency scattering constraints~\cite{Chen2026}. Building on this foundation, we explore the physics of generalized Huygens metasurfaces and formulate a framework for their rational design, extending the widely adopted Huygens platform beyond conventional dipole resonances to achieve high transmission efficiency and $2\pi$ phase excursion.

We note that this proposed metasurface design approach is distinct from conventional inverse-design schemes, which typically optimize phase and amplitude responses, or derived performance metrics, at real frequencies~\cite{Sell2017,Lin2018,Pestourie2018,Lin2019,Chung2020,Molesky2018,Jensen2011,Piggott2015}. Such schemes have been successfully applied to optimize both single- and multi-wavelength responses. Nevertheless, tailoring a resonant response introduces additional challenges. Specifying the response across a frequency band increases the number of sampled objectives or constraints; resonances with a high quality factor $Q$ require particularly fine or adaptively refined spectral sampling to resolve narrow features and rapid phase variations. More fundamentally, a direct response objective does not explicitly identify the resonances and interference mechanisms responsible for the target spectrum. Without an accompanying physical model, the optimization can therefore function as a black box: a favorable spectrum alone provides limited guidance for interpreting the solution, choosing an initial structure from physical heuristics or analytical solutions, or making rational design iterations. These limitations motivate a formulation that exposes the underlying resonance structure as a design input, rather than requiring it to be inferred from the optimized response.

To this end, we prescribe the locations of transmission and reflection singularities and optimize the underlying permittivity or geometry with adjoint derivatives. The pole coordinate sets the resonance center and linewidth, while channel-specific zeros govern interference and, together with the poles, the phase response. In particular, placing a transmission zero in the upper half plane while suppressing reflection near the real axis provides explicit targets for near-$2\pi$ phase control while maintaining high transmission~\cite{Colom2023}. This representation connects the desired optical response to physically interpretable design coordinates and provides common targets across different geometric representations. It also offers a direct way to incorporate resonance models into target selection and initialization. A matched-start single-resonance comparison in Sec.~3 illustrates linewidth and group-delay control, as well as lower FDFD work than direct transmission optimization.

We first establish these design controls using a binary guided-mode resonance grating, then apply the framework to the inverse design of several distinct classes of generalized Huygens metasurfaces. Continuous and binary high-index designs, a freeform low-index structure, passive and phase-change reconfigurable planar high-contrast grating (HCG) pairs, and a quasi-bound-state-in-the-continuum (qBIC) structure demonstrate how pole--zero targets can guide phase control across different dielectric contrasts, resonance mechanisms, geometric representations, and material selections. These demonstrations collectively establish a unified route to resonant phase control in both freeform and geometrically constrained metasurfaces.

\section{Inverse Design Method}
\subsection{Complex-frequency constraints and adjoint derivatives}
We adopt the time dependence $\exp(-i\omega t)$, where $\omega$ is angular frequency, $t$ is time, and $i$ is the imaginary unit. Pole and zero locations are specified using the complex vacuum wavenumber $k=\omega/c$, expressed in $\mu\mathrm{m}^{-1}$, where $c$ is the speed of light in vacuum. We consider normally incident light and metasurfaces whose periods are smaller than the wavelength in both input and output media throughout the spectral band of interest. Only the zeroth diffraction order then propagates on either side; for the polarization-preserving responses considered here, the metasurface is modeled as a two-port system. For identical uniform input and output media, the scattering matrix $S$ relates power-normalized incoming amplitudes $a_j$ to outgoing amplitudes $b_j$ at ports $j=1,2$:
\begin{equation}
\begin{pmatrix}b_1\\b_2\end{pmatrix}
=S(k)\begin{pmatrix}a_1\\a_2\end{pmatrix},
\qquad r(k)=S_{11}(k),\quad t(k)=S_{21}(k).
\label{eq:ports}
\end{equation}
Here $r(k)$ and $t(k)$ are the reflection and transmission amplitudes for incidence from port 1. The two-dimensional cells are periodic along $x$, invariant along $y$, and illuminated normally along $z$ with magnetic-field component $H_y$. On the real axis, the reflectance and transmittance are $R=|r|^2$ and $T=|t|^2$; for the ideal lossless model, $R+T=1$.

In a region without a diffraction branch point, a scalar channel can locally be written in the form
\begin{equation}
t(k)=B(k)\frac{\prod_m(k-z_m)}{\prod_n(k-p_n)},
\label{eq:factor}
\end{equation}
where $z_m$ and $p_n$ are the selected transmission zeros and poles, indexed by $m$ and $n$, respectively. The analytic background factor $B(k)$ is nonzero in that region and includes contributions not represented by the selected singularities. Passive resonance poles satisfy $\operatorname{Im}p_n<0$, where $\operatorname{Im}$ denotes the imaginary part.

The two-dimensional forward problem is discretized from
\begin{equation}
\left[-\nabla\cdot\epsilon_r^{-1}\nabla-k^2\right]H_y=J_{\mathrm{src}},
\label{eq:maxwell}
\end{equation}
where $\epsilon_r$ is relative permittivity, $\nabla$ acts in the $x$--$z$ plane, and $J_{\mathrm{src}}$ is the excitation source. We use a periodic finite-volume grid. At the upper and lower boundaries, an outgoing Rayleigh expansion represents the scattered field as diffraction waves. At real frequencies, these waves propagate or decay away from the structure; their outgoing branches are continued analytically to complex $k$. Their axial propagation constants use the grid's finite-difference dispersion relation, ensuring consistency between the boundary conditions and the interior discretization. The three-dimensional extension uses a full-vector Yee-grid frequency-domain operator, transverse periodicity, and axial perfectly matched layers. The design models use real, isotropic, nonmagnetic, and nondispersive permittivities.

Let $\mathbf{x}$ denote the vector of design variables. For a prescribed pole $p$ and outgoing coupling ratio $\sigma=D_2/D_1$, where $D_1$ and $D_2$ are the mode's radiation amplitudes into the two ports, time reversal gives the differentiable vector constraint~\cite{Chen2026}
\begin{equation}
S(p^*;\mathbf{x})
\begin{pmatrix}1\\\sigma^*\end{pmatrix}=\mathbf{0}.
\label{eq:chen}
\end{equation}
Here the superscript $*$ denotes complex conjugation. Physically, time reversal converts the outgoing resonant mode at $p$ into a purely incoming solution at $p^*$, whose incident wave amplitudes are the complex conjugates of the mode's outgoing amplitudes. In this lossless setting, the resulting exponentially growing excitation produces coherent virtual absorption: energy is stored in the structure rather than dissipated~\cite{Baranov2017}. In cells with mirror symmetry along the longitudinal (propagation) direction, the two opposite coupling parities reduce to $\sigma=\pm1$ after fixing the amplitude convention. Equation~\eqref{eq:chen} enforces the prescribed pole in the lossless time-reversal setting considered here. As required for each design, transmission and reflection zeros at prescribed coordinates $z_t$ and $z_r$ are imposed directly:
\begin{equation}
t(z_t;\mathbf{x})=0,\qquad
r(z_r;\mathbf{x})=0.
\label{eq:zeros}
\end{equation}
The selected channels must subsequently be checked for nonzero pole residues, since a mode can be invisible in a particular output. We also keep port reference planes fixed: changes in propagation reference alter phase backgrounds even when they do not move finite scalar zeros.

At a fixed complex wavenumber, the discretized field equation is $A(\mathbf{x})u=\mathbf{b}$, where $A$ is the Maxwell matrix, $u$ is the field vector, and $\mathbf{b}$ is the excitation vector. An outgoing amplitude is written as $q=Pu+d_{\mathrm{dir}}$, where $P$ projects the field onto the selected port and $d_{\mathrm{dir}}$ accounts for any direct contribution. Thus $S$ maps incoming port amplitudes to outgoing port amplitudes, whereas $P$ extracts one outgoing amplitude from the computed field: for unit excitation at port $j$ and projection onto port $i$, $S_{ij}=P_i u_j+d_{\mathrm{dir},ij}$, where $u_j$ is the field for that excitation and $d_{\mathrm{dir},ij}$ is its direct contribution to port $i$. Depending on the excitation and output port, $q$ represents $t$, $r$, or an outgoing amplitude under the prescribed two-port excitation in Eq.~\eqref{eq:chen}. With the excitation and port definitions held fixed, the adjoint field $\psi$ satisfies $A^T\psi=P^T$, giving the derivative with respect to design variable $x_j$:
\begin{equation}
\frac{\partial q}{\partial x_j}=-\psi^T
\frac{\partial A}{\partial x_j}u.
\label{eq:adjoint}
\end{equation}
The superscript $T$ denotes the ordinary transpose, rather than the Hermitian transpose, as required for this derivative of the complex amplitude. Real and imaginary parts are supplied as separate rows to the optimizer. The derivative of $A$ is evaluated by the chain rule, accounting for both the permittivity averaging at grid-cell faces and the mapping from design variables to the material distribution. For geometric variables, the same adjoint expression accounts for the change in the material distribution caused by interface motion~\cite{LalauKeraly2013}.

For the pole constraint in Eq.~\eqref{eq:chen}, we evaluate the field at $k=p^*$ with simultaneous illumination from both ports, using $\mathbf{b}=\mathbf{b}^{(1)}+\sigma^*\mathbf{b}^{(2)}$, where $\mathbf{b}^{(1)}$ and $\mathbf{b}^{(2)}$ are the excitation vectors for unit incidence from ports 1 and 2. The outgoing amplitudes are $q_i=S_{i1}(p^*;\mathbf{x})+\sigma^*S_{i2}(p^*;\mathbf{x})$ for $i=1,2$, and the pole constraint requires $q_1=q_2=0$. Equation~\eqref{eq:adjoint} applies separately to each amplitude, using the corresponding port projection $P_i$ and adjoint field $\psi_i$, with the prescribed $p$ and $\sigma$ held fixed. Both derivatives use the same forward field $u$.

The reflection channel is included in this formulation by choosing $q=r$. We seek $r(z_r)=0$ with $z_r$ close to the real axis and, where necessary, reduce reflection at several real-frequency samples surrounding it. The design rules in Sec.~\ref{sec:kerker-rules} explain how this reflection constraint complements transmission pole--zero targeting.

\subsection{Material representation and continuation}
Continuous topology variables are represented by a density field $\rho$, whose value in grid cell $j$ satisfies $0\leq\rho_j\leq1$. To suppress fine-scale variations in density-based designs, we apply a Helmholtz filter with radius $r_f$, $(I-r_f^2\nabla^2)\rho_f=\rho$, where $I$ is the identity operator and $\rho_f$ is the filtered density; setting $r_f=0$ recovers the unfiltered density. A projection then gives the projected density $\rho_p$ and material permittivity~\cite{Wang2011Projection}:
\begin{equation}
\rho_p=\frac{\tanh(\beta\eta)+\tanh[\beta(\rho_f-\eta)]}
{\tanh(\beta\eta)+\tanh[\beta(1-\eta)]},\qquad
\epsilon_r=[n_l+(n_h-n_l)\rho_p]^2.
\label{eq:density}
\end{equation}
Here $\beta$ and $\eta$ set the projection sharpness and threshold, respectively, and $n_l$ and $n_h$ are the low and high refractive indices. The filter and projection parameters are configuration-specific. To obtain a two-material binary structure from a continuous-index design, we increase $\beta$ in stages, progressively driving the projected density toward 0 or 1 on either side of $\rho_f=\eta$. At each stage, the density variables are reoptimized against the scattering constraints, starting from the preceding solution. This allows the material boundaries to adjust as intermediate refractive indices are removed, rather than simply thresholding the original continuous-index design. The finite-sharpness projection remains differentiable, so the adjoint gradients propagate through the projection and filter to the density variables. After continuation, hard thresholding defines a fixed two-material geometry whose poles, zeros, and optical response are checked independently. That same geometry is represented on each validation mesh using subpixel averaging at interfaces; intermediate numerical permittivities then describe partial cell filling. The continuation procedure and design-specific settings are detailed in SI, Sec.~S4.

The staged projection sharpening described above is one form of continuation: a sequence of related optimization problems in which each solution initializes the next stage. We also use continuation to move singularity targets from an accessible starting state toward their desired values before adding channel-zero or background-reflection conditions. Changes to the geometric representation are handled in separate stages, with independent verification after each representation change. Within each continuation stage, the scattering targets and the current material or geometric representation, including any filter and projection settings, are held fixed while the design variables are optimized. At each iteration, forward solves evaluate the mismatch with the prescribed scattering constraints, and adjoint solves provide its derivatives with respect to the design variables. The weighted complex mismatches are split into real and imaginary parts to form a real residual vector $\mathbf{f}$ with Jacobian $J=\partial\mathbf{f}/\partial\mathbf{x}$. The density problems use a bounded damped least-squares update in the underdetermined Levenberg--Marquardt form~\cite{Chen2026}; when variables outnumber residuals, the implementation solves
\begin{equation}
\left(JJ^T+\mu_d\frac{\|J\|_F^2}{N_{\mathrm{var}}}I\right)\mathbf{v}
=-\mathbf{f},\qquad \delta\mathbf{x}=J^T\mathbf{v},
\label{eq:lm}
\end{equation}
where $N_{\mathrm{var}}$ is the number of design variables, $\mu_d$ is the damping parameter, $\|J\|_F$ is the Frobenius norm of the Jacobian, $I$ is the identity matrix, and $\mathbf{v}$ is an auxiliary vector used to obtain the design step $\delta\mathbf{x}$. Steps are restricted to the admissible bounds, and damping is adjusted according to residual reduction. Few-coordinate problems use bounded least-squares formulations in their smaller parameter space.

Details of independent root refinement, mesh validation, and phase span evaluation are given in SI, Sec.~S5.

\section{Single-resonance design: physically interpretable control}
\label{sec:single}
As discussed above, prescribing complex pole locations directly specifies resonance frequencies and linewidths and provides a physically transparent route to controlling group delay. Encoding these resonance properties in a small number of complex-frequency constraints can reduce the number of spectral constraints required by conventional real-frequency spectrum matching and lower computational cost. In this section, we use a single-resonance guided-mode-resonance (GMR) structure to validate these design capabilities and quantify their computational advantages relative to direct transmission optimization.

\subsection{Guided-mode-resonance grating benchmark}
We consider a continuous dielectric slab with one centered rectangular ridge per period, both with refractive index 2.0 and surrounded by air. The structure is invariant along $y$ and illuminated normally by linearly polarized light with a magnetic field $H_y$; the ridge occupies 66\% of the period. The period sets the lateral phase-matching condition, while slab thickness and ridge height control modal confinement and radiative coupling. This binary, geometrically parameterized structure supports a transmission notch associated with a leaky GMR.

We characterize each resonance by its complex pole $p=k_r-i\gamma$, where $k_r=\operatorname{Re}p$ is the resonance-center wavenumber and $\gamma=-\operatorname{Im}p>0$ is the pole depth. The quality factor is defined as $Q=k_r/(2\gamma)$. This pole-based definition avoids the ambiguity of assigning a full width at half maximum to the asymmetric Fano transmission line shape~\cite{Fan2003}. We prescribe target poles $p_{\mathrm{tar}}=k_0-i\gamma_{\mathrm{tar}}$ with a common center $k_0=6.77~\mu\mathrm{m}^{-1}$ (vacuum wavelength approximately 928~nm) and depths $\gamma_{\mathrm{tar}}=k_0/(2Q_{\mathrm{tar}})$ for target quality factors $Q_{\mathrm{tar}}=50,100,200,400,800,1600,$ and 3200.

For each target, a preliminary Fourier phase-matching search selects the grating period. With this design-specific period held fixed, the finite-difference frequency-domain (FDFD) adjoint optimizes slab thickness and ridge height by imposing $\det S(p_{\mathrm{tar}}^*;\mathbf{x})=0$, leaving the outgoing coupling ratio unspecified. The final physical geometries are held fixed for verification with inverse-factorized TM rigorous coupled-wave analysis (RCWA). For each final geometry, a complex-wavenumber root search locates a zero of $1/t(k)$ near the prescribed pole; the refined pole determines the verified $Q$. The exemplar spectra in Fig.~\ref{fig:gmr-combined}a show progressively narrower resonances near the common target wavelength. The ridge height decreases from 182.6 to 32.1~nm across the family, consistent with weaker radiative coupling at higher $Q$. Geometry and numerical implementation details are provided in SI, Sec.~S6.

\subsection{Linewidth and group-delay control}
The prescribed pole depth directly specifies the full resonance linewidth in ordinary frequency, $\Delta\nu_{\mathrm{tar}}=c\gamma_{\mathrm{tar}}/\pi$, where $\nu=\omega/(2\pi)$. The verified linewidth is evaluated from the corrected RCWA pole as $\Delta\nu=c\gamma/\pi$. Figure~\ref{fig:gmr-combined}b shows agreement between the target and verified linewidths across all seven frozen geometries. The verified linewidths are larger than their targets by 3.4--8.7\%. These deviations include FDFD discretization and the original outer-correction model; no geometry was reoptimized after the RCWA formulation correction.

To relate this linewidth control to transmission group delay, write the local response as $t(k)=C(k)/(k-p)$, where $C(k)$ contains the transmission zeros and nonresonant background. The group delay $\tau=d\arg t/d\omega$ is
\begin{equation}
\tau(k)=\frac{1}{c}\operatorname{Im}\!\left[\frac{d\log C}{dk}-\frac{1}{k-p}\right].
\label{eq:delay}
\end{equation}
The pole contributes $\gamma/\{c[(k-k_r)^2+\gamma^2]\}$, linking its decay rate to an explicit delay scale. We evaluate the full RCWA group delay at $k_{\mathrm{op}}=k_r-2\gamma$, away from the transmission notch. At this common normalized detuning, the prescribed pole implies a target delay $\tau_{\mathrm{tar}}=1/(5c\gamma_{\mathrm{tar}})$.

Figure~\ref{fig:gmr-combined}c compares this pole-implied target with the verified transmission group delay. The relative deviations between the target and verified delays are at most 5.0\% for $Q_{\mathrm{tar}}\geq100$ and 14.3\% for $Q_{\mathrm{tar}}=50$. The nonresonant background that is omitted from the isolated-pole prediction contributes to such deviations; the background has a larger relative influence at low $Q$, where the pole-induced delay is smaller. These results demonstrate control of the group-delay scale through the prescribed linewidth.

\subsection{Computational cost comparison with direct transmission optimization}
We quantify computational cost through recovery of two known feasible GMR reference geometries, corresponding to $Q_{\mathrm{tar}}=200$ and 800. For each reference, pole targeting minimizes the real and imaginary parts of $\det S(p_{\mathrm{ref}}^*;\mathbf{x})$, where $p_{\mathrm{ref}}$ is its independently extracted FDFD transmission pole. For the selected resonance, the scattering-matrix and transmission poles coincide. The direct alternatives instead match three or five complex transmission amplitudes at real frequencies near the resonance. Their targets are calculated by full-wave simulation of the same reference geometry. Thus both approaches seek a physically attainable reference response, while differing in how that response is specified.

All three methods optimize slab thickness and ridge height using adjoint derivatives and the same bounded least-squares solver, material indices, period, geometric bounds, and FDFD grid. Each method is tested from the same twelve perturbed geometries; the signed starting offsets are also matched between the two references. Each start has a budget of 300 objective-frequency factorizations. We count the sparse-matrix factorizations needed for all attempts, including objective evaluations, target generation and updates, spectral recentering, and transmission pole extraction after each optimization stage. These pole searches are benchmark verification, not part of the direct-transmission objective. Shared reference characterization is charged once to each method. This accounting measures the dominant FDFD computational work rather than elapsed time. Protocol and accounting details are given in SI, Sec.~S6.

\begin{figure}[p]
\centering
\includegraphics[width=\linewidth,height=0.62\textheight,keepaspectratio]{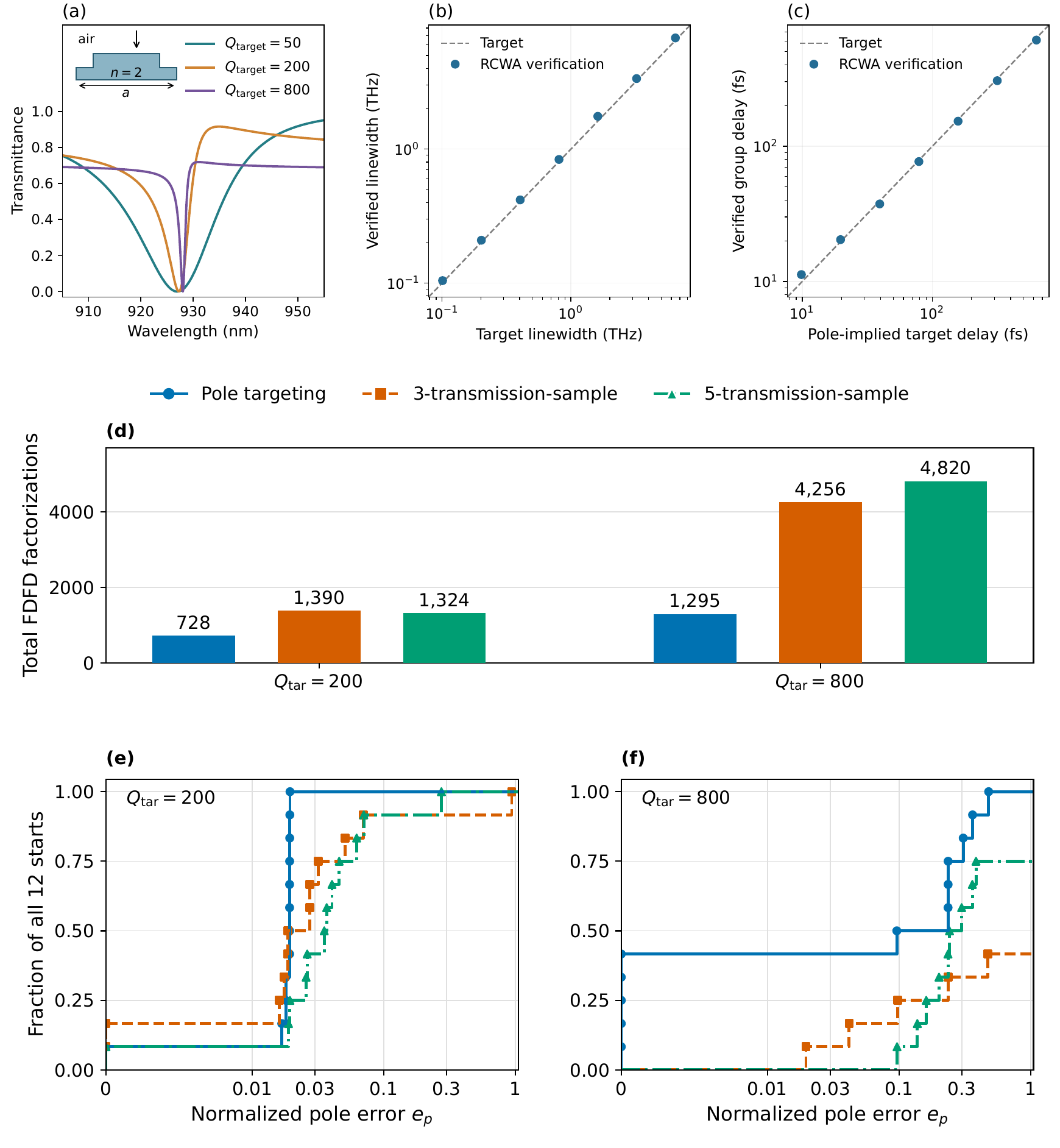}
\small
\caption{Pole-based specification of resonance parameters and computational-cost comparison in binary GMR structures. (a) Simulated transmission spectra for grating structures with target Q-factors $Q_{\mathrm{tar}}=50,200,$ and 800. Inset: one period $a$ of the ridge-on-slab grating in air, with refractive index $n=2$. (b) Verified resonance linewidth $\Delta\nu$ from pole-based inverse design versus prescribed linewidth $\Delta\nu_{\mathrm{tar}}$. (c) Verified transmission group delay $\tau$ at $k_{\mathrm{op}}=k_r-2\gamma$ from inverse design versus the pole-implied target $\tau_{\mathrm{tar}}$. (d--f) Computational work and pole recovery for twelve matched starts per method, comparing pole targeting with direct matching of three or five real-frequency complex transmission samples. (d) Total FDFD factorization counts, including all attempts, target preparation and updates, recentering, pole extraction, and shared reference characterization. (e,f) Fraction of the twelve optimization runs yielding a normalized complex-pole error $e_p$ no greater than the value on the horizontal axis for $Q_{\mathrm{tar}}=200$ and 800. Curves closer to the upper-left corner indicate more accurate recovery of the target pole. The horizontal scale is linear up to 0.01 and logarithmic above it. Curves that remain below unity include outputs for which no reliable nearby pole was resolved: seven for three-sample and three for five-sample matching at $Q_{\mathrm{tar}}=800$.}
\label{fig:gmr-combined}
\end{figure}

Pole targeting requires 728 total FDFD factorizations for the $Q_{\mathrm{tar}}=200$ reference, compared with 1390 and 1324 for three- and five-sample transmission matching, respectively (Fig.~\ref{fig:gmr-combined}d). For the $Q_{\mathrm{tar}}=800$ reference, the corresponding counts are 1295, 4256, and 4820. Pole targeting therefore uses approximately 1.8--1.9 times fewer counted factorizations for the lower-$Q$ reference and 3.3--3.7 times fewer for the higher-$Q$ reference in this protocol. The objective-only counts, excluding target preparation, recentering, stage-end pole verification, and shared reference characterization, are 581, 1029, and 1020 for $Q_{\mathrm{tar}}=200$, and 1120, 3129, and 3600 for $Q_{\mathrm{tar}}=800$, in the same method order.

To quantify how closely the optimized designs approach the target pole, we evaluate the normalized complex-pole error
\begin{equation}
e_p=\frac{|p_{\mathrm{out}}-p_{\mathrm{ref}}|}{\gamma_{\mathrm{ref}}},
\qquad \gamma_{\mathrm{ref}}=-\operatorname{Im}p_{\mathrm{ref}},
\end{equation}
where $p_{\mathrm{out}}$ is extracted independently from the optimized geometry's transmission response. Panels (e--f) of Fig.~\ref{fig:gmr-combined} show the fraction of all twelve starts attaining each error level. For $Q_{\mathrm{tar}}=200$, nearby poles are resolved for every output, and median errors are 0.019, 0.023, and 0.036 for pole targeting, three-sample matching, and five-sample matching, respectively. For $Q_{\mathrm{tar}}=800$, pole targeting again yields resolved nearby poles for all twelve starts, with median error 0.166; the direct methods yield resolved nearby poles for five and nine starts, respectively. These results demonstrate reduced computational load and more reliable recovery of the targeted resonance branch in this benchmark. Independent RCWA checks and application-specific qualification criteria, including cost per qualified design, are reported in the SI.

\section{Resonant metasurface Designs}
While Sec.~3 establishes the capabilities and advantages of the pole--zero inverse-design framework, this section applies it to a central challenge in resonant metasurface design: engineering resonances to realize a generalized Kerker response, combining suppressed reflection and near-unity transmission with a nearly $2\pi$ transmission-phase excursion. We first summarize the resonance-control criteria that enable this response, then use these physical insights to guide the inverse design of a diverse set of metasurfaces. We begin with two-dimensional freeform structures, demonstrating that the framework accommodates both continuous-index and binary material distributions, including high- and low-index designs. We then extend the approach beyond conventional Mie-resonant meta-atoms to structures supporting other resonance mechanisms, using HCGs and qBICs (presented in the SI) as examples. Finally, we apply the framework to reconfigurable phase-change metasurfaces, where the design must maintain the Kerker response across a continuum of refractive-index states rather than at a single material configuration. These examples demonstrate how explicit pole--zero targets provide a common, physically interpretable design strategy across geometric representations, resonance mechanisms, and passive and active material platforms.

\subsection{Pole--zero design rules for the generalized Kerker condition}
\label{sec:kerker-rules}
For a reciprocal metasurface with mirror symmetry along the propagation direction and identical input and output media, a two-resonance route to the generalized Kerker condition can be formulated in terms of complex poles, longitudinal parity, and scattering residues. Under a consistent port convention, the symmetric and antisymmetric scattering eigenchannels are $s_\pm=r\pm t$. Near a pair of simple resonances, write
\begin{equation}
s_\pm(k)=B_\pm(k)+\frac{\mathcal R_\pm}{k-p_\pm},
\qquad
r=\frac{s_++s_-}{2},\quad t=\frac{s_+-s_-}{2},
\label{eq:parity-residues}
\end{equation}
where $p_\pm$ are the eigenchannel poles, $\mathcal R_\pm$ are their complex residues and $B_\pm$ contain the nonresonant background and contributions from other modes. This representation yields three design rules:
\begin{enumerate}
\item \textit{Match the complex poles.} Set $p_+=p_-=k_0-i\gamma$. Matching the real and imaginary parts simultaneously aligns the resonance centers and decay rates. In particular, the pole linewidth $2\gamma$ and quality factor $Q=k_0/(2\gamma)$ are already matched by this single complex equality; linewidth matching is not an additional constraint. This is an advantage over direct real-frequency transmission/reflection matching, which generally requires multiple spectral samples to constrain both resonance position and width. The pole linewidth refers to the isolated Lorentzian contribution; interference with zeros and background can modify the observed lineshape.
\item \textit{Select opposite longitudinal parity.} The two resonances should couple to opposite port-parity channels, corresponding to symmetric and antisymmetric radiation under exchange of the two sides of the metasurface. Their contributions can then cancel in reflection while adding in transmission. This requirement concerns radiation symmetry, not a restriction to electric and magnetic dipoles, and therefore accommodates guided and collective resonances.
\item \textit{Balance the complex residues and background.} For matched poles $p=p_+=p_-$, the resonant reflection is $(\mathcal R_++\mathcal R_-)/[2(k-p)]$. Its cancellation requires $\mathcal R_+=-\mathcal R_-$: equal residue magnitudes and a relative phase of $\pi$ in this convention. The transmitted resonant contributions then add. Total reflection also includes $(B_++B_-)/2$, which must be negligible or compensated over the desired band. Pole matching alone therefore does not guarantee reflection suppression; reflection-zero constraints enforce the coherent cancellation of the full response, including its background.
\end{enumerate}
These rules translate the familiar balance of electric and magnetic responses into a scattering-channel description~\cite{Decker2015,Colom2023,Binkowski2024}, and provide a constructive recipe for a two-resonance Huygens-like response. For a prescribed center wavenumber $k_0$ and quality factor $Q_{\mathrm{tar}}$, the two-resonance optimization targets can be written explicitly as
\begin{equation}
\begin{aligned}
p_{\mathrm{tar}}&=k_0-i\gamma_{\mathrm{tar}},
\qquad \gamma_{\mathrm{tar}}=\frac{k_0}{2Q_{\mathrm{tar}}},\\
s_+(p_{\mathrm{tar}}^*;\mathbf{x})&=0,
\qquad s_-(p_{\mathrm{tar}}^*;\mathbf{x})=0,\\
r(z_{r,j};\mathbf{x})&=0,\qquad j=1,\ldots,N_{r,0}.
\end{aligned}
\label{eq:kerker-targets}
\end{equation}
Here $\gamma_{\mathrm{tar}}$ is the prescribed pole depth, and $z_{r,j}$ are the $N_{r,0}$ selected reflection-zero targets. The first two scattering constraints are the opposite-parity forms of Eq.~\eqref{eq:chen}: they prescribe a common outgoing pole at $p_{\mathrm{tar}}$, with full wavenumber linewidth $2\gamma_{\mathrm{tar}}=k_0/Q_{\mathrm{tar}}$. The reflection constraints control cancellation of the full response; real targets $z_{r,j}=k_j$ enforce $R(k_j)=0$ and, in the lossless two-port model, $T(k_j)=1$ at the selected operating wavenumbers $k_j$. The real and imaginary parts of these complex amplitudes enter the adjoint least-squares objective.

\subsection{2D freeform high-index Huygens surface}
We begin with high-index dielectric Huygens surfaces, a classical platform for realizing the Kerker condition through interference between electric and magnetic resonances. Early demonstrations employed simple meta-atom geometries, such as silicon nanodisks, whose well-defined electric and magnetic Mie-type dipole resonances could be brought into spectral overlap by adjusting their dimensions~\cite{Decker2015,Yu2015,Zhao2016}. Here, we instead explore freeform geometries with spatially varying refractive index, allowing more complex, non-intuitive resonant field distributions without restricting the design to a predefined pair of Mie modes.

While these earlier studies establish the principle of reflection cancellation, near-unity transmission at resonance does not ensure uniformly low reflection across the spectral interval supporting a near-$2\pi$ phase excursion. Residual reflection can remain at the several-percent level within this band, even when cancellation is nearly complete at resonance~\cite{Decker2015}. Suppressing reflection throughout the band requires coordinating the resonance linewidths, coupling strengths, and nonresonant response as well as the resonance frequencies. We therefore use this example to test whether matching the complex poles while prescribing reflection zeros on the real axis can significantly reduce the maximum residual reflection across a nearly $2\pi$ phase excursion. We then examine whether the continuous-index freeform design retains this response when driven toward a binary, two-material structure.

We first consider a cell with period $a=0.36~\mu\mathrm{m}$, maximum index $n_h=3.7$, and background index $n_l=1.45$. The starting structure comprises two axially coupled bars, each $0.28~\mu\mathrm{m}$ wide and $0.16~\mu\mathrm{m}$ thick, separated by a $0.16~\mu\mathrm{m}$ spacer. Freeform density optimization allows both bars and the surrounding design region to change. Opposite-parity poles are continued toward the common target $p=7.90802-0.324982i~\mu\mathrm{m}^{-1}$, followed by reflection constraints.

\begin{figure}[p]
\centering
\includegraphics[width=\linewidth]{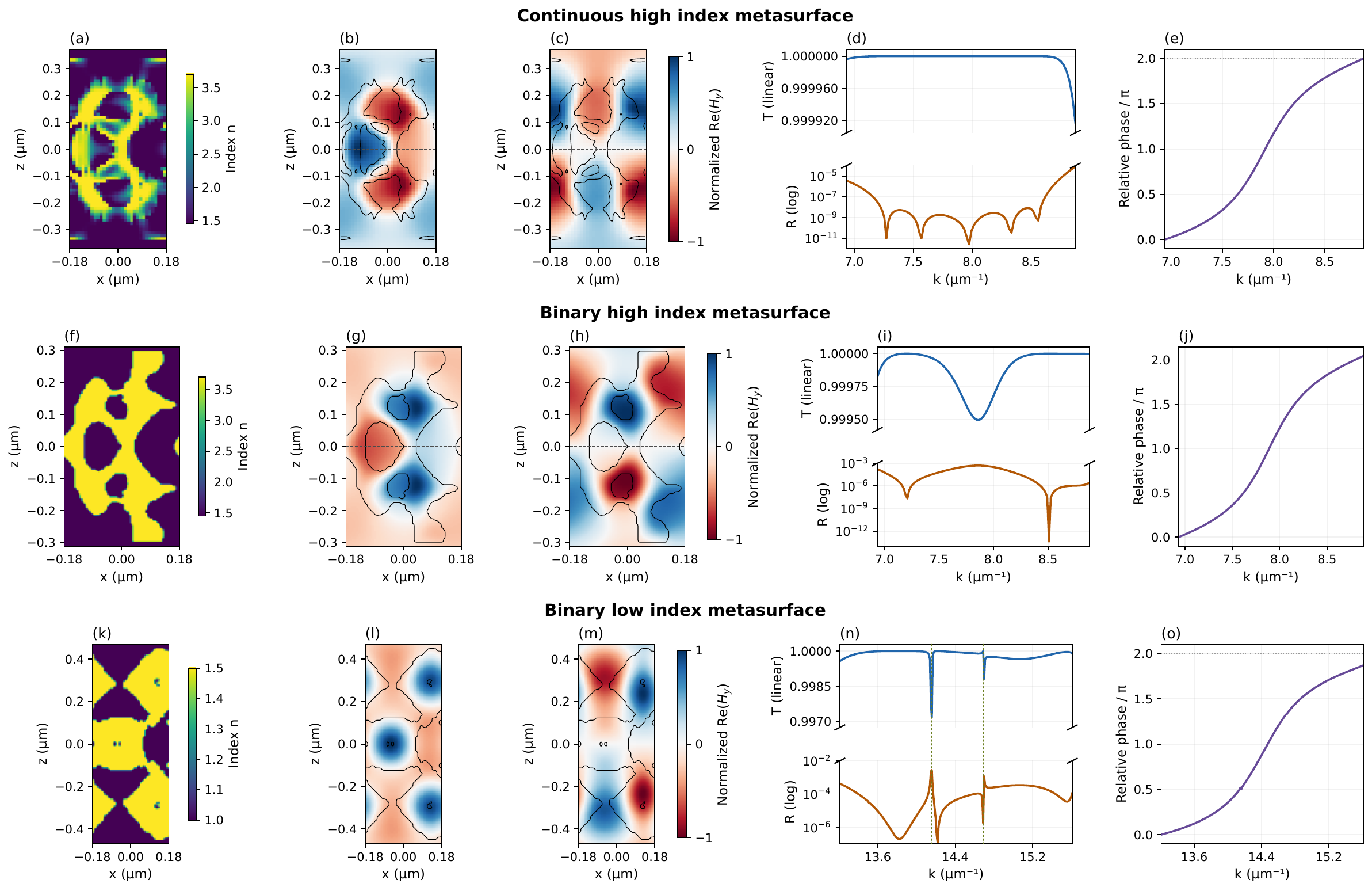}
\caption{Freeform 2D metasurface designs. Rows show continuous high-index, binary high-index, and binary low-index designs. Columns show refractive index (a, f, k); longitudinally symmetric (b, g, l) and antisymmetric (c, h, m) resonant-mode profiles, shown as normalized $\operatorname{Re}(H_y)$; real-frequency transmittance and reflectance (d, i, n); and unwrapped transmission phase (e, j, o), with its first plotted value subtracted. Each complex mode is independently normalized to its peak magnitude and phase-aligned to make its largest-magnitude sample in the upper half of the cell positive real; the maps show its signed real part. Black contours mark the midpoint of the permittivity range, and dashed lines mark the longitudinal symmetry plane. The binary low-index response contains two in-band high-$Q$ reflection peaks marked by olive dotted guides. The coarse spectra in (n) do not resolve these fine features; pole-guided local high-resolution sweeps are shown in SI. Intermediate interface colors in binary profiles arise from subpixel averaging. Geometry and field panels retain equal physical scales along $x$ and $z$.}
\label{fig:freeform}
\end{figure}

For the continuous-index design (Fig.~\ref{fig:freeform}a), independent scalar refinement gives a transmission pole $7.908000-0.325020i~\mu\mathrm{m}^{-1}$ and an upper-half-plane transmission zero $7.908020+0.324982i~\mu\mathrm{m}^{-1}$. Reflection-channel root searches place a near-real-axis zero around $k=7.968~\mu\mathrm{m}^{-1}$; the last quoted digits depend on root-refinement settings (SI, Sec.~S5). Across $6.93307\leq k\leq8.88297~\mu\mathrm{m}^{-1}$, corresponding to three target pole depths on either side of the center, the plotted samples give $T_{\min}=0.999917$ and $R_{\max}=8.29\times10^{-5}$, together with nearly $2\pi$ transmission-phase variation across the resonance (Fig.~\ref{fig:freeform}d--e). This continuous-index design therefore demonstrates high sampled transmission and nearly full-cycle phase control.

Projection continuation and binary reoptimization produce a fixed two-material contour (Fig.~\ref{fig:freeform}f) with $T_{\min}=0.999496$ and $R_{\max}=0.000504$ over the same band, while retaining nearly $2\pi$ transmission-phase variation across the resonance (Fig.~\ref{fig:freeform}i--j). The binarization and numerical validation are detailed in SI, Secs.~S4 and S5.

The resonant-mode profiles in Fig.~\ref{fig:freeform}b--h reveal the opposite longitudinal parities of the paired resonances. The even and odd $H_y$ modes are magnetic-like and electric-like in their normal-incidence radiation symmetry, respectively. However, transverse harmonic analysis confirms that these modes are lattice-guided in nature and are not described by the isolated-particle dipole model commonly used for classical nanodisk Huygens metasurfaces. This design thus exemplifies a \emph{generalized} Huygens surface.

\subsection{2D freeform low-index Huygens surface}
We next extend the freeform \emph{generalized} Huygens metasurface design to low-index-contrast systems. Low-index materials already enable successful phase-control metasurfaces, including silica platforms that accumulate propagation phase in waveguide-type meta-atoms~\cite{Sacchi2025,Ueno2026}. The challenge here is different: to obtain low reflection and a near-$2\pi$ spectral phase excursion through overlapping resonances, rather than propagation phase. Although low-index spheres have demonstrated Huygens-type directional scattering~\cite{Abdelrahman2019}, this does not establish the corresponding transmission-phase response of a metasurface. Here we test this resonance-overlap strategy at an index comparable to those of glasses and polymers (around 1.5).

To address this gap, we use pole matching and reflection-zero constraints to seek overlapping resonances with suppressed reflection in a freeform dielectric structure with refractive index $n_h=1.5$ in air. We retain the $0.36~\mu\mathrm{m}$ period of the high-index examples but expand the axial design region, giving the optimizer greater geometric freedom to realize the required resonance pair at low index contrast. Starting from a continuous-density precursor, we tune the resonances toward a target quality factor $Q=18$ and then reoptimize the design as a connected binary structure. The binarization procedure protects a periodic material backbone, suppresses detached islands, and retains a shared master representation of the material boundary.

Over a frequency range spanning $13.20705\leq k\leq15.60834~\mu\mathrm{m}^{-1}$, the binary profile gives a $1.8694\pi$ transmission-phase range (Fig.~\ref{fig:freeform}n--o). Unlike the previous example, however, the spectrum contains two high-$Q$ in-band reflection peaks near $k=14.15315$ and $14.69437~\mu\mathrm{m}^{-1}$; both approach unit reflectance (SI Fig.~S2). Outside the two narrow dips, $T_{\min}=0.997180$ and $R_{\max}=0.002820$ (Fig.~\ref{fig:freeform}n). Thus the structure demonstrates broadband reflection suppression via opposite-parity mode overlap, but \emph{not} uninterrupted high transmission throughout this band. That said, the two resolved intervals with $T<0.9$ occupy only approximately $0.08\%$ of the studied bandwidth; in practice, they could be avoided by selecting wavelengths away from the peaks. The mode maps in Fig.~\ref{fig:freeform}l--m identify the targeted resonances; fixed-geometry mesh trends and further analysis of the narrow features are reported in SI, Sec.~S5.

\subsection{2D high-contrast grating (HCG)}
Having established resonance control in freeform metasurfaces, we next examine whether the same framework can realize the Kerker condition in a simpler geometry amenable to standard lithography and etching. A pair of planar high-contrast gratings provides such an example: its rectangular ridges are described by a small set of widths, heights, and separations, without requiring a freeform final geometry. We therefore use this structure to test whether pole--zero constraints can guide the optimization of geometrically parameterized interfaces while retaining high transmission and nearly $2\pi$ resonant phase variation.

\begin{figure}[tb]
\centering
\includegraphics[width=\linewidth]{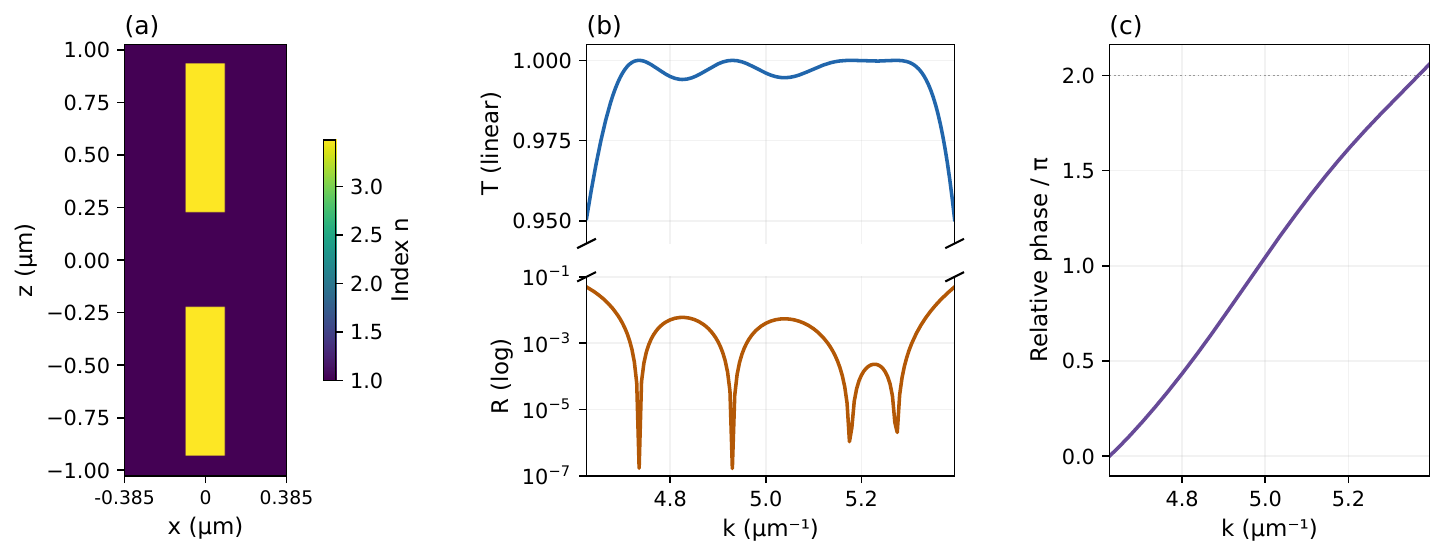}
\caption{Planar binary HCG: (a) index profile, (b) transmittance and reflectance, and (c) unwrapped transmission phase. The geometry panel shows one period with equal physical scales along $x$ and $z$.}
\label{fig:hcg}
\end{figure}

For this design, we fix the cell period to be $0.77~\mu\mathrm{m}$ and grating index to be 3.48 in air. The optimized structure consists of two identical rectangular gratings, each $0.192500~\mu\mathrm{m}$ wide and $0.705833~\mu\mathrm{m}$ thick, separated by an air gap of $0.449167~\mu\mathrm{m}$ (Fig.~\ref{fig:hcg}a). Across $4.625\leq k\leq5.395~\mu\mathrm{m}^{-1}$, the transmittance remains well above 90\%, with $T_{\min}=0.950361$ and $R_{\max}=0.049639$, while the transmission phase varies by nearly $2\pi$ across the resonance (Fig.~\ref{fig:hcg}b--c). This result extends the framework from freeform structures to a compact, geometrically parameterized design. The optimization targets, geometric continuation, and numerical checks are detailed in SI, Sec.~S5.

The framework has also been implemented for a three-dimensional metasurface recovery test (SI, Sec.~S2) with the present limits on quantitative mesh convergence stated in SI, Sec.~S7.

\subsection{Phase-change reconfigurable metasurface}
\label{sec:pcm}
We now examine a reconfigurable design. Generalized Huygens metasurfaces provide an attractive platform for reconfigurable meta-optics based on tunable materials. Their ultrathin profiles facilitate integration with active thin films, whose unique tunable properties are often inaccessible in bulk form~\cite{Zhang2021PCM,Wuttig2017}. The small active volume can reduce switching energy and accelerate tuning relative to bulk or discrete optical components. Furthermore, high-$Q$ resonances can translate relatively small changes in material optical constants into large changes in the optical response, enabling a substantial phase-tuning range without requiring a comparably large refractive index change. These opportunities motivate a different design specification from the passive examples above: maintaining high transmission while tuning the phase at a fixed wavelength across multiple material states with varying optical constants. We demonstrate that the pole--zero inverse-design framework can accommodate this multistate objective.

As a demonstration, we replace both gratings by the phase-change material Sb$_2$Se$_3$. The design model uses the reported 1550-nm amorphous- and crystalline-state endpoint indices $n_a=3.285$ and $n_c=4.050$~\cite{Delaney2020}, approximates both states as lossless and isotropic, and interpolates their permittivities using Lorentz--Lorenz mixing to obtain the optical constants of intermediate states~\cite{Xu2026PCM}, which are assumed to be spatially homogeneous.

\begin{figure}[p]
\centering
\includegraphics[width=\linewidth]{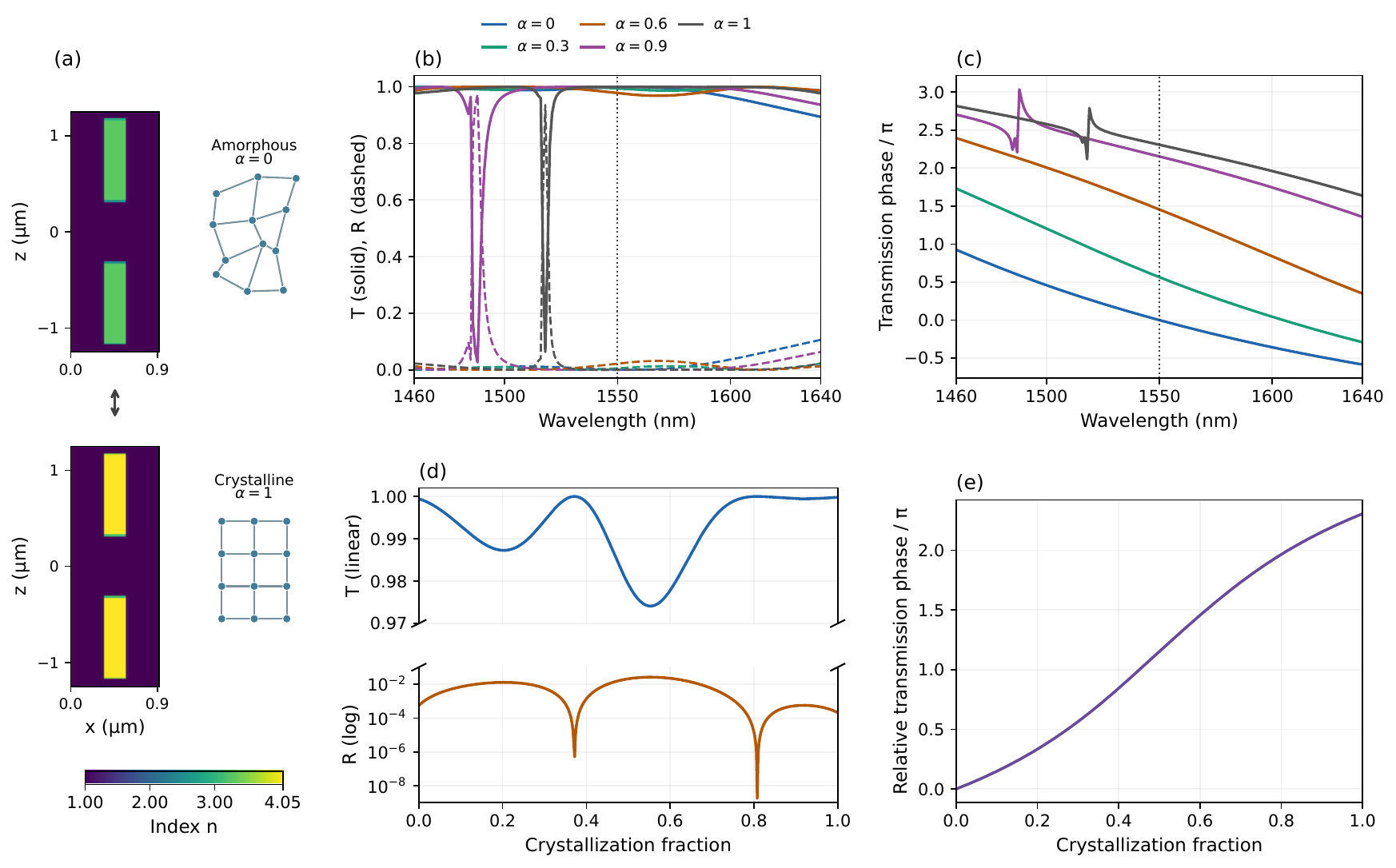}
\caption{Phase-change Sb$_2$Se$_3$ paired-HCG design. (a) Amorphous-state ($\alpha=0$, top) and crystalline-state ($\alpha=1$, bottom) index profiles on a common color scale, with equal physical scales. (b) Transmittance (solid) and reflectance (dashed) spectra at crystallization fractions $\alpha=0,0.3,0.6,0.9,$ and 1. (c) Transmission-phase spectra for the same states, using the colors in b. Vertical dotted lines in b and c mark the operation wavelength 1550~nm. (d) Transmittance and reflectance at 1550~nm across different crystallization fractions. (e) Corresponding optical phase tuning relative to the amorphous state.}
\label{fig:pcm}
\end{figure}

The objective combines an even-parity pole target with reflection-channel zeros at 1550~nm for crystallization fractions $\alpha=0,0.3,0.6,0.9,$ and 1. No transmission or phase samples enter the objective. The optimized structure has a period of $0.921739~\mu\mathrm{m}$ and two identical rectangular gratings, each $0.851166~\mu\mathrm{m}$ tall and $0.231171~\mu\mathrm{m}$ wide, separated by a $0.631201~\mu\mathrm{m}$ air gap (Fig.~\ref{fig:pcm}a).

The spectra in Fig.~\ref{fig:pcm}b--c show how changes in crystallization fraction reshape the transmission, reflection, and phase responses, with the operating wavelength of 1550~nm marked by vertical dotted lines. At 1550~nm, a 501-state sweep gives $T_{\min}=0.974111$ and an unwrapped phase range of $2.3061\pi$ over $0\leq\alpha\leq1$ (Fig.~\ref{fig:pcm}d--e). Thus the multistate reflection-zero constraints support high transmission throughout the continuum of material states without directly fitting transmission phase. Optimization details are reported in SI, Sec.~S3.

\clearpage
\section{Fabrication-aware optimization}
\label{sec:fabrication-aware}
\begin{figure}[tb]
\centering
\includegraphics[width=\linewidth]{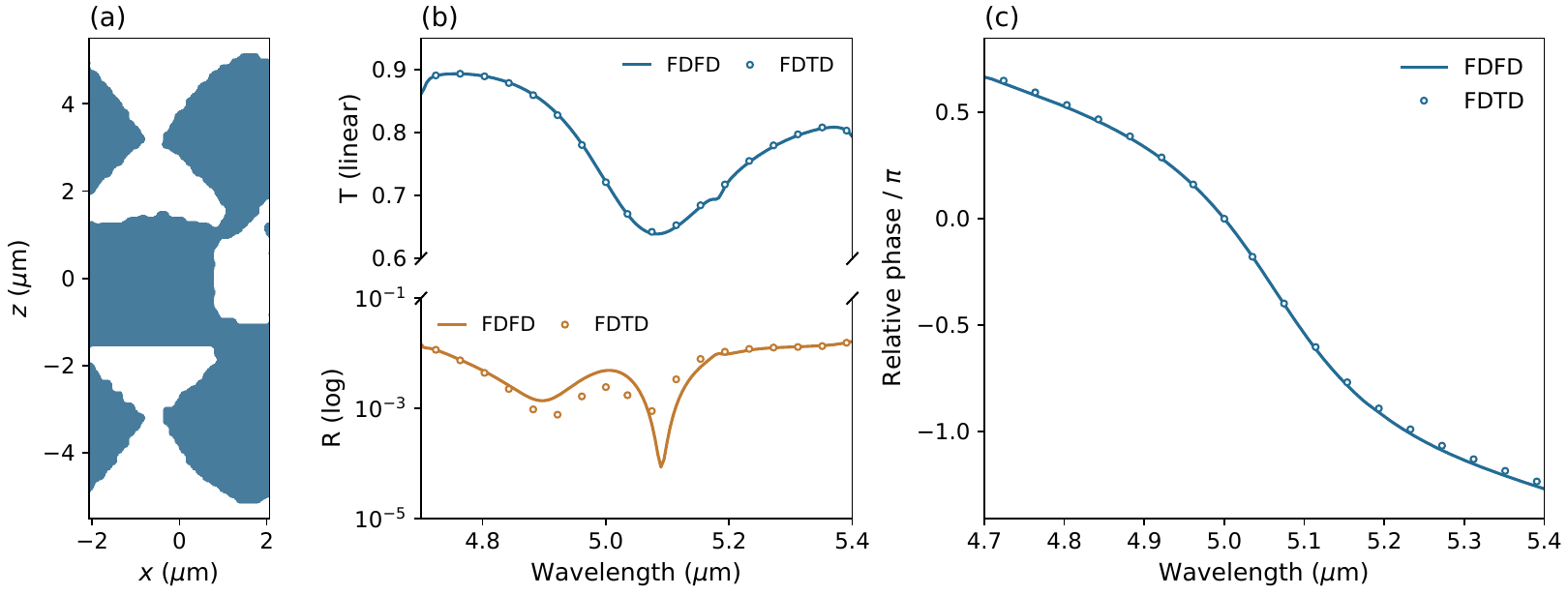}
\caption{Fabrication-aware connected polymer metasurface under TM-polarized illumination. (a) One period of the final binary polymer contour, with air elsewhere; the axes have equal physical scales. (b) Transmittance and reflectance calculated using the measured dispersive, absorptive material response. (c) Transmission phase after removing empty-cell propagation simulated using FDFD and FDTD, referenced to its value at $5~\mu\mathrm{m}$.}
\label{fig:fabrication-aware}
\end{figure}

The preceding examples establish resonance control across different material and geometric platforms. Translating this capability into a realizable metasurface also requires consideration of fabrication constraints and the optical properties of the actual material. We therefore consider a connected IP-Dip polymer metasurface for fabrication via two-photon polymerization, designed for TM illumination around $5~\mu\mathrm{m}$. The design uses measured, wavelength-dependent complex refractive indices~\cite{Fullager2017IPDip}; at $5~\mu\mathrm{m}$, the index is $1.50555+0.004815i$. In addition to the optical constraints, the optimization imposes a 200-nm minimum feature size in both polymer and air and requires all parts of the structure to form a connected network as a safeguard against buckling. The resulting structure has a period of $4.1275~\mu\mathrm{m}$ and an axial extent of $10.319~\mu\mathrm{m}$ (Fig.~\ref{fig:fabrication-aware}a).

Fabrication constraints are enforced during optimization rather than applied only to the final optical design~\cite{Piggott2017Fabrication}. Adjoint derivatives rank candidate boundary modifications, which are projected onto an explicitly binary contour satisfying the minimum feature constraints. Each candidate must retain a connected dielectric network across neighboring periods and keep air voids open to the exterior. The optical response is recalculated on the projected geometry before an update is accepted. The final contour satisfies the 200-nm polymer and air feature tests; a local clearance analysis places its smallest solid connecting feature at approximately $0.30$--$0.31~\mu\mathrm{m}$ and its narrowest air gap for developer access at approximately $0.26~\mu\mathrm{m}$ (SI, Sec.~S8).

These geometric requirements and the polymer optical absorption change the design trade-offs relative to the idealized low-index example in Sec.~4.3. Material connections introduce additional resonances, while absorption limits transmission and makes long resonant dwell times undesirable. We therefore combine complex-frequency pole constraints with sparse real-frequency reflection and absorptance controls, and favor boundary updates that increase transmission at $5~\mu\mathrm{m}$. Deeper pole targets encourage broader resonances, providing a physical route to reduce the sensitivity to absorptive loss without directly fitting the transmission phase. For the absorptive material, the time-reversed pole constraint conjugates the material response as well as the frequency, as detailed in the SI.

The final structure exhibits near-complete reflection cancellation and nearly $2\pi$ of transmission-phase variation across $4.7$ -- $5.4~\mu\mathrm{m}$ (Fig.~\ref{fig:fabrication-aware}b--c). FDFD gives a maximum reflectance of 1.65\% over this band and, at $5~\mu\mathrm{m}$, $T=0.722$ and $R=0.00486$. The corresponding absorptance, $A=1-T-R$, is 0.273: material absorption, rather than reflection, is the main limitation on transmission. Independent FDTD calculations validate the spectral and phase responses. Thus the framework accommodates measured material dispersion and loss together with connectivity and minimum feature constraints, retaining broad reflection suppression and a nearly full-cycle phase response in a fabrication-aware design.

\section{Discussion and conclusion}
We have implemented an inverse-design framework that turns complex-frequency poles and channel zeros from descriptors of resonant scattering into explicit design targets for metasurface phase control. Building on adjoint pole-based design, the framework combines resonance placement with transmission- and reflection-channel constraints to engineer \emph{generalized} Huygens responses. Its central advantage is the direct connection between the optimization targets and the underlying physics: complex poles specify resonance frequencies and decay rates, while channel zeros control destructive interference. In particular, matching complex poles incorporates frequency and linewidth matching in one condition, providing a physically interpretable route to the Kerker response without restricting the design to conventional electric and magnetic dipole geometries.

The single-resonance benchmark demonstrates how this representation enables rational control of linewidth and group delay with a compact set of scattering constraints. In the matched-start optimization comparison, pole targeting reduces the total counted FDFD factorizations, including stage-end pole verification, by approximately 1.8--1.9 times for the lower-$Q$ reference and 3.3--3.7 times for the higher-$Q$ reference relative to direct transmission matching, while improving recovery of the targeted resonance branch. The objective-only counts show the same ordering. These results illustrate a computational benefit of specifying the resonance itself rather than resolving its spectral signature through multiple real-frequency samples in this benchmark. More broadly, physically meaningful targets provide guidance for selecting initial structures, interpreting optimization outcomes, and deciding which resonance properties to modify in subsequent design iterations.

The metasurface examples establish the versatility of this approach across refractive index contrasts, resonance mechanisms, and geometric representations. High-index freeform structures and planar gratings demonstrate broadband low reflection with nearly full-cycle spectral phase control. The low-index freeform and qBIC studies also expose important limits: unresolved narrow spectral features interrupt the former's high transmission, and a continuous qBIC endpoint does not guarantee a recoverable binary design. The three-dimensional Huygens meta-atom example presented in the SI provides a further physical test. The phase-change example extends the design constraints across material states, maintaining high transmission while enabling more than a full cycle of phase tuning at a fixed wavelength. Fabrication-aware optimization shows how the same framework can accommodate connected binary structures and explicit minimum-feature constraints. Using measured optical constants, the connected design retains low reflection and nearly full-cycle phase variation despite appreciable absorption, illustrating how material loss and fabrication constraints can be treated jointly.

Looking ahead, joint control of resonant phase, linewidth, and delay could support dispersive metalenses, compact pulse-shaping elements, and wavelength-selective wavefront control. Extending the material-state formulation to spatially addressable tunable elements could enable reconfigurable beam steering and adaptive meta-optics~\cite{Popescu2026SLM}. Realizing these functions will require translating the demonstrated spectral and material-state responses into spatial phase profiles while accounting for coupling between nonidentical cells. The same design philosophy could also be extended to multiple polarizations, incidence angles, and diffraction channels, with channel-specific zeros used to select the desired scattering pathways. Pole--zero inverse design thus offers a common physical language for linking resonance engineering, computational optimization, and practical metasurface functionality.

\section{Methods}
\paragraph{Computing environment.}
The in-house calculations and post-processing were implemented in Python 3.12 with NumPy, SciPy, and Matplotlib. 2-D FDFD simulations were executed locally, except for the fine-grid GMR computational-cost benchmark, which was performed using Amazon Web Services (AWS). 3-D FDFD simulations were also executed on AWS. OpenAI Codex was used for code development and to execute Python scripts in the local computing environment.

\paragraph{FDFD solvers.}
The 2-D and 3-D FDFD solvers were developed in Python using NumPy and SciPy, independently of the commercial solver used for validation. The 2-D FDFD solver discretizes the scalar TM equation for $H_y$ on a cell-centered grid, with harmonic averaging of the material coefficients at cell faces, transverse periodic boundary conditions, and outgoing-wave conditions in uniform axial ports. The 3-D solver uses a full-vector curl--curl formulation on a Yee grid, periodic boundaries in the two transverse directions, and perfectly matched layers along the propagation direction. Sparse linear systems are solved using SciPy's direct sparse solvers.

\paragraph{FDTD validation.}
2-D and 3-D FDTD validation uses the commercial Tidy3D solver (version 2.12.0, Flexcompute). Device simulations are normalized using reference simulations without the metasurface, with identical source, monitor, and boundary settings. Both use periodic transverse boundaries and axial absorbing boundaries: perfectly matched layers for the examples in Sec.~4 and an adiabatic absorber for the IP-Dip structure. SI, Secs.~S7 and S8 specify the validation meshes, spectral sampling, normalization, and finite-duration checks.

\paragraph{Code and data availability.}
The Python source, numerical configurations, final geometries, selected frozen verification spectra, and the complete 72-trial fine-grid benchmark record are available at \url{https://github.com/hujuejun/metasurface-pole-zero-design}. Portable Tidy3D simulation definitions and the numerical comparison snapshots are also supplied.

\printbibliography

\end{document}